\documentclass[a4paper, amsfonts, amssymb, amsmath, reprint, showkeys, nofootinbib, superscriptaddress, floatfix]{revtex4-2}
\usepackage[english]{babel}
\usepackage[utf8]{inputenc}
\usepackage{amsthm}
\usepackage{mathtools}
\usepackage{physics}
\usepackage{xcolor}
\usepackage{graphicx}
\usepackage[left=23mm,right=13mm,top=35mm,columnsep=15pt]{geometry} 
\usepackage{adjustbox}
\usepackage{placeins}
\usepackage[T1]{fontenc}
\usepackage{lipsum}
\usepackage{csquotes}
\usepackage{tikz}
\usepackage{float}
\usetikzlibrary{quantikz}
\usepackage[unicode=true, colorlinks=true]{hyperref} 

\usepackage[caption=false]{subfig}

\usepackage{multirow} 
\usepackage{soul} 
\usepackage{packages/nicematrix}

\newcommand{\mynorm}[1]{\left\lVert#1\right\rVert}

\usepackage{packages/orcidlink}

\newcommand{\UC}{Yusuf Hamied Department of Chemistry, University of Cambridge\\ Lensfield Road, CB2 1EW Cambridge\\ United Kingdom}
\newcommand{\QT}{Quantinuum\\ Terrington House, 13-15 Hills Road, CB2 1NL Cambridge\\ United Kingdom}

\begin{document}
\title{Non-unitary Trotter circuits for imaginary time evolution}

\author{Chiara Leadbeater\,\orcidlink{0000-0001-6577-3926}}
    \email{cnl29@cam.ac.uk}
    \affiliation{\UC}
    \affiliation{\QT}
\author{Nathan Fitzpatrick\,\orcidlink{0000-0001-5819-9129}}
    \email{nathan.fitzpatrick@quantinuum.com}
    \affiliation{\QT}
\author{David \surname{Mu\~{n}oz Ramo}\,\orcidlink{0000-0002-7220-9466}}
    \email{david.munozramo@quantinuum.com}
    \affiliation{\QT}
\author{Alex J. W. Thom\,\orcidlink{0000-0002-2417-7869}}
    \email{ajwt3@cam.ac.uk}
    \affiliation{\UC}

\date{\today} 

\begin{abstract}

We propose an imaginary time equivalent of the well-established Pauli gadget primitive for Trotter-decomposed real time evolution, using mid-circuit measurements on a single ancilla qubit. Imaginary time evolution (ITE) is widely used for obtaining the ground state of a system on classical hardware, computing thermal averages, and as a component of quantum algorithms that perform non-unitary evolution. Near-term implementations on quantum hardware rely on heuristics, compromising their accuracy. As a result, there is growing interest in the development of more natively quantum algorithms. Since it is not possible to implement a non-unitary gate deterministically, we resort to the implementation of probabilistic imaginary time evolution (PITE) algorithms, which rely on a unitary quantum circuit to simulate a block encoding of the ITE operator -- that is, they rely on successful ancillary measurements to evolve the system non-unitarily. Compared with previous PITE proposals, the suggested block encoding in this paper results in shorter circuits and is simpler to implement, requiring only a slight modification of the Pauli gadget primitive. This scheme was tested on the transverse Ising model and the fermionic Hubbard model and is demonstrated to converge to the ground state of the system.
\end{abstract}

\maketitle

\section{Introduction} \label{sec:introduction}

Imaginary time evolution (ITE) is a routine that is widely used on classical hardware to obtain the ground state (GS) of a system. In this procedure, an initial trial state is evolved through imaginary time, $\tau = it$, by applying the ITE operator, $e^{-\hat{H}\tau}$. This operator has the property of being a ground state projector in the large $\tau$ limit; over time, the excited state contributions from the initial state $\ket{\Psi(\tau=0)}$ decay until only the GS, $\ket{\phi_0}$, remains
\begin{equation}
    \lim_{\tau \rightarrow \infty} e^{-(\hat{H}-E_0)\tau} \ket{\Psi(\tau=0)} \propto \ket{\phi_0}. \label{eq:ITE_operator}
\end{equation}
Although it is not expected that quantum computers will be able to make the task of ground state determination for generic Hamiltonians efficient \cite{kempe2006complexity}, they may provide useful speedups in the near future for certain applications using quantum dynamics \cite{miessen2023dynamicsrev}. In particular, owing to the exponential scaling of the Hilbert space with system size, classical techniques for quantum simulation are generally designed to avoid two features in a computation: (1) explicitly storing the many-body wave function, and (2) propagating the wave function by matrix exponentiation. Quantum computers allow for efficient implementation of both of these features. Regarding the former, there are quantum circuits which cannot be efficiently simulated using classical resources \cite{lund2017quantumsupremacy}. The quantum computer thus has access to a richer state space for the ground state search of certain systems. Regarding the latter feature, there are quantum algorithms for efficiently simulating real time evolution. Each algorithm corresponds to a different decomposition of the real time evolution operator; for instance, the operator can be expressed as a truncated Taylor series, which can be implemented via a linear combination of unitaries (LCU) \cite{berry2015lcu, childs2012lcu, berry2014exponential, berry2015simulation}, or it can be expressed as a series of alternating signal rotations and signal processing rotations via quantum signal processing (QSP) \cite{low2017qsp}. Due to its simplicity and ease of implementation with the Pauli gadget primitive (refer to Appendix \ref{appendix:appendix_pauli_gadget}) \cite{nielsenchuang2011}, the most widely adopted method is the Trotter decomposition (alternatively referred to as the `product formula' approach) \cite{lloyd1996universalsimulators, nielsen2002universal, aharanov2003adiabatic, jones2018trotterisation}, which evolves each Hamiltonian term for a small time slice. \newline

Imaginary time evolution is much more challenging to implement, owing to the non-unitarity of the ITE operator: since quantum gates are unitary operations, the ITE operator cannot be implemented deterministically on quantum hardware. Many of the previous approaches to this problem have focussed on implementing the ITE operator indirectly using a hybrid quantum--classical approach. These algorithms are thus designed for hardware in the current `noisy intermediate-scale quantum' (NISQ) era, characterised by a limited coherence time and gate fidelity. Whilst the partitioning between classical and quantum hardware saves quantum resources, it typically necessitates the use of heuristics, which can hinder the attainable accuracy of the final ground state. Indeed, the variational quantum eigensolver \cite{peruzzo2014vqe, mcclean2016vqe}, which is among the most promising and widely-used examples of variational algorithms, requires gate errors significantly lower than the error thresholds of fault-tolerant hardware in order to achieve chemical accuracy \cite{dalton2022ftqc}. Within the class of hybrid ITE techniques, two main approaches have emerged: variational ITE (VITE) \cite{jones2019vite, mcardle2019vite, yuan2019vite, amaro2022jobscheduling, benedetti2021hardwareefficient} and quantum ITE (QITE) \cite{motta2020determining, gomes2020qite, sun2021qite, nishi2021implementation}. Refer to Section \ref{sec:hybrid_methods} for more information regarding these methods. \newline



Unlike their near-term competitors, fault-tolerant algorithms do not rely on heuristics and offer rigorous performance guarantees. To this end, we consider a third approach: the probabilistic ITE (PITE). PITE algorithms are a type of block encoding -- that is, the non-unitary operator is embedded inside a larger unitary matrix, which is accessed via post-selection of `successful' mid-circuit measurements. \newline

Unlike VITE, PITE does not require an Ansatz, and unlike QITE, it does not require the selection of Pauli strings according to the locality of the Hamiltonian. PITE evolves the system exclusively using quantum operations, without the need for classical updates to the state. The main drawback of PITE algorithms arises from the need to implement them as a series of iterative circuits \cite{silva2021fragmentedQITE}, each of which must be applied successfully -- that is, each imaginary time step is implemented using at least one successful mid-circuit measurement. Consequently, as with most block encoding methods, PITE algorithms are characterised by an exponentially decaying probability of success; though they are guaranteed to converge to the GS, convergence is limited by the number of measurement shots required. The practicality of implementing most block encoding methods, including PITE, will be tied to the development of amplitude amplification (AA) schemes to boost the overall success probability of the circuits \cite{grover1996fast, grover1997haystack, dalzell2017fixedpoint}. AA has been utilised to boost PITE circuits with some success \cite{nishi2022accelerating}.\newline



The ITE operator can be decomposed using similar strategies as the RTE operator; namely, QSVT \cite{suri2022twounitary, chan2023simulating}, LCU and Trotterisation. PITE algorithms use mid-circuit measurements, which naturally renormalise the quantum state throughout the evolution process; this represents an advantage of PITE methods over other approaches which must compute this normalisation, including hybrid methods such as QITE, as well as QSVT. In recent years, a number of PITE algorithms have been  proposed \cite{liu2021pite, zhang2021observation, silva2021fragmentedQITE, kosugi2021probabilistic} and implemented on hardware \cite{xie2022pite, turro2022itehardware}. Amongst these, the LCU proposal from Kosugi et al \cite{kosugi2021probabilistic} is the most comparable to the scheme that will be proposed in this paper: both methods are general schemes for constructing the necessary circuits directly given any input Hamiltonian, without the need for system-specific block encodings and without incurring an additional classical cost in computing an initial singular value decomposition of the ITE operator \cite{liu2021pite}. In the LCU proposal, a first-order approximation of the ITE operator, $(1-\hat{H}\tfrac{\tau}{r})^r$, where $r$ is the number of time steps, is achieved by the repeated application of controlled forward and backward RTE operators, requiring two controlled Pauli gadgets for each Hamiltonian term per time step. \newline

In this paper, we propose an alternative block encoding of the ITE operator, assuming a Trotter-decomposed form. The algorithm presented can be considered to be the imaginary time counterpart of the well-established Trotterised RTE algorithm, which makes use of the Pauli gadget primitive. There are several advantages to using this approach: firstly, the structure of the Pauli gadget naturally results in diagonal matrices, avoiding the need to incur an initial classical cost in computing the singular value decomposition (SVD) of the ITE operator \cite{liu2021pite}. Secondly, the similarity of the proposed Trotterised ITE algorithm to the Trotterised RTE algorithm also presents advantages. Trotterisation is an active area of research, owing to its simplicity and unique scaling with the commutative structure of the Hamiltonian. In RTE simulations, this can be leveraged to achieve much better empirical performance than more sophisticated algorithms, including LCU-Taylor and QSP \cite{childs2018speedup}. By building on the Pauli gadget framework, we ensure that any of the optimisation passes previously designed for RTE Pauli gadgets can be directly transferred to our circuits \cite{cowtan2020phase}. Further to this, the Trotter decomposition is highly flexible, with many flavours extensively reviewed by the community; as such, there are many options readily available for improving the accuracy of the evolution, which is of particular importance for the task of thermal state preparation -- on the other hand, the LCU PITE proposal is only a first-order approximation of the ITE operator. Finally, it is easy to account for non-Hermitian Hamiltonians by concatenating the proposed (probabilistic) imaginary time Pauli gadgets with the ubiquitous (deterministic) real time Pauli gadgets. \newline

The structure of the paper is as follows. Background information is given in Section \ref{sec:background}, describing the implementation of non-unitary operators on quantum circuits. The proposed algorithm will then be presented in Section \ref{sec:method}, and the results obtained will be detailed in Section \ref{sec:results}. Finally, a discussion of the results will be presented in Section \ref{sec:discussion}, and conclusions drawn in Section \ref{sec:conclusion}. \newline

\section{Background} \label{sec:background}

\subsection{NISQ ITE Methods} \label{sec:hybrid_methods}

Heuristic ITE approaches approximate the action of the non-unitary ITE propagator with a unitary operator. This unitary operator is optimised by taking measurements and post-processing the results classically. Within this class of techniques, two main approaches have emerged: variational ITE (VITE) \cite{jones2019vite, mcardle2019vite, yuan2019vite, amaro2022jobscheduling, benedetti2021hardwareefficient} and quantum ITE (QITE) \cite{motta2020determining, gomes2020qite, sun2021qite, nishi2021implementation}. \newline

VITE represents the unitary operator with a parameterised quantum circuit, whose parameters are optimised classically according to a variational principle \cite{mcardle2019vite} which carries out a finite projection of the wavefunction on the tangent space of the exact evolution. The standard VITE optimisation procedure requires the inversion of a matrix whose elements are determined by measurements made on the quantum circuit; due to the sensitivity of matrix inversion to external environmental pertubations, this procedure assumes very low gate errors \cite{saxena2023practical}. Matrix inversion can be avoided, and thus the hardware requirements of the standard VITE procedure can be significantly reduced, by assuming a Trotter product form for the Ansatz and variationally optimising each Trotter term sequentially \cite{benedetti2021hardwareefficient}. Further, optimising the circuits using classical tensor network methods can result in shallower and more accurate circuits \cite{keever2022classically}. Variational quantum algorithms (VQAs) such as these have substantial drawbacks \cite{tilly2021vqe}, including the barren plateau (BP) problem, which could prevent convergence for larger problem sizes \cite{mcClean2018bps}. There are multiple driving factors; in particular, the more expressive the Ansatz is, the more susceptible it is to BPs \cite{mcClean2018bps, holmes2021bps}. A common approach to mitigate the BP problem is to reduce the expressibility of the Ansatz. In the context of quantum chemistry problems, this corresponds to restricting the span of the Ansatz to a section of the Hilbert space, for instance by exploiting the symmetry of the problem \cite{holmes2021bps}. Consequently, VQAs are highly dependent on the choice of Ansatz. \newline


On the other hand, QITE represents the unitary operator in a Pauli basis and solves a linear system of equations for the coefficients of the Pauli tensors; measurements taken from the quantum circuit are used to estimate the matrix elements, and the equations are solved on classical hardware. The derivation is mathematically equivalent to McLachlan's variational principle for VITE \cite{kosugi2021probabilistic}. However, the domain of these Pauli tensors grows exponentially with the spreading of entanglement. As a result, the matrix dimension for the linear equations, and correspondingly the number of measurements, tends to grow rapidly as the problem size increases. Its scaling can be regularised if restrictive approximations regarding the locality of Hamiltonian terms are adopted \cite{nishi2021implementation}.

\subsection{Block encoding scheme for non-unitary operation} \label{sec:block_encoding}

Block encoding represents a general framework for the implementation of non-unitary operators: a non-unitary matrix $A$ that operates on an $n$-qubit quantum state $\ket{\psi}$ is embedded as a block inside a larger unitary matrix $U_A$ that operates on $(n+l)$ qubits. In theory, any non-unitary matrix A can be block encoded with a single ancilla using a singular value decomposition scheme \cite{lin2022lecture}. However, these circuits may be exponentially deep.  Application of $A$ is then contingent on successfully post-selecting the $l$ ancillary qubits corresponding to this block, $\ket{0}^{\otimes l}\bra{0}^{\otimes l} \otimes \ket{\psi} \bra{\psi}$ of the unitary matrix $U_A$. For the purpose of this work, we will iteratively use a single ancilla qubit, i.e. $l=1$, block encoding. When $l=1$, the block encoded matrix is given by
\begin{equation}\label{eq:block_encoding_matrix}
    U_A = \begin{pNiceMatrix}[first-row, last-col]
    & \ket{0} & \ket{1} \\
& \tfrac{1}{\alpha}A & * & \ket{0} \\
& * & * & \ket{1}\\
\end{pNiceMatrix} \;\;,
\end{equation}
where $\tfrac{1}{\alpha} \in \mathbb{R}$ is a necessary scaling required to ensure that $U_A$ is unitary; more precisely, the spectral norm of $\tfrac{1}{\alpha} A$ satisfies $\|\tfrac{1}{\alpha}A\| = \tfrac{1}{|\alpha|}\|A\| \leq 1$. The spectral norm $\| \cdot \|$ of a matrix $A$ is the maximum scale by which $A$ can stretch a vector, defined as
\begin{align}
    \|A\| &= \sup_{\langle x|x\rangle = 1} \sqrt{\bra{x} A^\dag A \ket{x}} \label{eq:spectral_norm_defn_op_norm}\\
    &= \sigma_{\text{max}} \left( A\right) \label{eq:spectral_norm_defn_sing_value},
\end{align}
where $\sigma_{\text{max}} \left( A\right)$ is the maximum singular value of $A$. Entries given by an asterisk $*$ indicate that there is some freedom in choosing their values -- again, the only requirement is that $U_A$ is unitary. The action of $U_A$ is:
\begin{equation} \label{eq:action_of_block_encoding}
    \ket{\Psi} = U_A (\ket{0}\otimes \ket{\psi}) = \ket{0} \otimes \tfrac{1}{\alpha}A \ket{\psi} + \ket{\perp}.
\end{equation}
The first and second terms represent the components of the state $\ket{\Psi}$ that are parallel and perpendicular to $\ket{0}$ in the auxilliary space, respectively. That is, $\ket{\perp}$ must satisfy the following orthonormality constraints:  $(\bra{0} \otimes I^{\otimes n})\ket{\perp} = 0$ and $\bra{\perp}\ket{\perp} = \sqrt{1 - \tfrac{1}{\alpha^2} \bra{0}\ket{0} \bra{\psi}A^\dag  A\ket{\psi}}$, but its system space component is determined by the action of the bottom left element in $U_A$ \eqref{eq:block_encoding_matrix} on $\ket{\psi}$. To successfully apply $A$, one must measure the ancilla qubit to be in the state $\ket{0}$. \newline



According to the postulates of quantum mechanics, following a partial measurement on the ancillary register, the remaining state vector is renormalised, preserving its purity. Thus, the value of $\tfrac{1}{\alpha}$ does not affect the successful post-measurement state. To see this, consider the projective measurement operator $\hat{M}_0 = \left(\ket{0}\bra{0} \otimes I^{\otimes n} \right)$, which only acts on the ancilla qubit, applied to the state $\ket{\Psi}$. The successful ancillary measurement outcome is defined to be

\begin{align}
     \begin{split}
     \ket{\Psi}
     &\rightarrow \frac{\hat{M}_0 \ket{\Psi}}{\sqrt{\bra{\Psi}\hat{M}_0^\dag \hat{M}_0\ket{\Psi}}} \\
     &= \frac{\ket{0} \otimes \tfrac{1}{\alpha}A \ket{\psi}}{\sqrt{\tfrac{1}{\alpha^2}\bra{0}\ket{0} \bra{\psi}  A^\dag A \ket{\psi}}} \\
    &= \frac{\ket{0} \otimes A \ket{\psi}}{\sqrt{\bra{\psi}A^\dag A\ket{\psi}}}.
    \end{split}
    \label{eq:renormalisation}
\end{align}


The value of $\tfrac{1}{\alpha}$ does, however, affect the probability of successfully applying $A$,
\begin{equation} \label{eq:success_probability}
    p_{\text{success}} = \tfrac{1}{\alpha^2} \bra{\psi} A^\dag A \ket{\psi}.
\end{equation}

The dependence of the success probability on the initial state $\ket{\psi}$ is a consequence of the non-unitarity of $\tfrac{1}{\alpha}A$ -- roughly speaking, the more non-unitary $\tfrac{1}{\alpha}A$ is, the more information that is lost from $\ket{\psi}$ on application of $\tfrac{1}{\alpha}A$, and the smaller the success probability is. Of course, the non-unitarity of $\tfrac{1}{\alpha} A$ is affected by the value of the scaling, $\tfrac{1}{\alpha}$: the greater the scaling, the greater the success probability. Different block encodings $U_A$ of $A$ may produce different scalings; however, there is a maximum value this scaling can take, $\left(\tfrac{1}{\alpha}\right)_{\text{max}}$. Above this value, the block encoding $U_A$ is no longer unitary and cannot be implemented with a quantum circuit. This optimal block encoding -- that is, the block encoding that applies $A$ with the maximal success probability for any $\ket{\psi}$ -- satisfies $\|\left(\tfrac{1}{\alpha}\right)_{\text{max}}A\|=1$, giving
\begin{equation}\label{eq:max_alpha}
    \left( \frac{1}{\alpha} \right)_{\text{max}} = \frac{1}{\|A\|} = \frac{1}{\sigma_{\text{max}} \left(A \right)}.
\end{equation}
\newline

When the scaling of a block encoding is not yet optimal, a classically-controlled reversing measurement scheme may be applied to improve the success probability \cite{terashima2005nonunitary}. In this scheme, an unsuccessful measurement -- that is, the application of the projective measurement operator $\hat{M}_1 = \ket{1}\bra{1} \otimes I^{\otimes n}$ on the state $\ket{\Psi}$ -- triggers the application of a reversing non-unitary operator, $\hat{R}_0$, defined such that $\hat{R}_0 \hat{M}_1 \ket{\Psi} \propto \ket{\Psi}$. The more times the reversing non-unitary is applied, the more the success probability is improved; in the limit of an infinite number of applications of $\hat{R}_0$, the optimal success probability is reached, 
\begin{equation} \label{eq:success_probability_max}
    p_{\text{success, max}} = \frac{1}{\|A\|^2} \bra{\psi} A^\dag A \ket{\psi}.
\end{equation}
\newline

The intuition behind the block encoding approach can be seen by drawing an analogy with the physical process of system-environment interactions. In physical systems, non-Hermicity is generated from entanglement with the environment; in Equation \eqref{eq:action_of_block_encoding}, we can identify the `system' to be the $n$-qubit state $\ket{\psi}$, whereas the `environment' is represented by the ancillary qubit. First, $U_A$ entangles the system with the environment. Next, the local measurement on the ancilla state, before checking the measurement result, produces a mixed state and corresponds to the process of taking the partial trace over the environmental degrees of freedom, resulting in a non-unitary evolution of the system. In particular, however, we are interested in the non-unitary evolution of the component of the system state vector that is entangled with the $\ket{0}$ ancilla state, $\ket{//}$. Initially, we begin in the normalised state $\ket{//} = \ket{0} \otimes \ket{\psi}$. Application of $U_A$ is a continuous real time evolution process, throughout which the magnitude of $\ket{//}$ decreases. At the end of this process, we have $\ket{//} = \ket{0} \otimes \tfrac{1}{\alpha}A\ket{\psi}$ and the magnitude of $\ket{//}$ has decreased by a factor of $\sqrt{p_{\text{success}}}$ (Equations \eqref{eq:action_of_block_encoding} and \eqref{eq:success_probability}). Finally, after looking at the measurement result, the system collapses into the corresponding pure state. This is the second process: discontinuous renormalisation of the state vector (Equation \eqref{eq:renormalisation}). \newline

\subsection{Universal gate set for non-unitary operation with a single ancillary qubit}\label{sec:measurement_scheme}

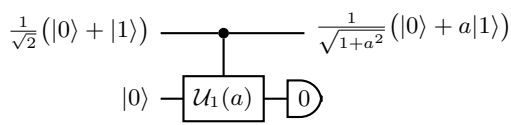
\begin{figure}[h]
\centering
\sbox0{
\begin{quantikz}[align equals at=1.5,column sep=0.3cm]
    \lstick[wires=1]{$\frac{1}{\sqrt{2}}\big(|0\rangle + |1\rangle \big)$} & \ctrl{1} & \qw \rstick[wires=1] {$\frac{1}{\sqrt{1+a^2}}\big(|0\rangle + a|1\rangle \big)$} \\
    \lstick[wires=1]{$\ket{0}$} & \gate{\mathcal{U}_1(a)} & \meterD{0}
\end{quantikz}
}
\begin{tabular}{c}
\usebox0
\end{tabular}
\caption{The non-unitary nature of $\mathcal{N}_1(a)$ projecting out the magnitude of $|0\rangle$ relative to $|1\rangle$ upon successful measurement of $|0\rangle$ on the ancilla.
}
\label{fig:N1}
\end{figure}

The {\sc cnot} gate and all single qubit unitary gates constitute a universal set for the unitary quantum circuit. Terashima et al. \cite{terashima2005nonunitary} proved that, considering only block encodings with one ancillary qubit, $l=1$, a possible universal set for non-unitary circuits is given by the {\sc cnot} gate, all single qubit unitary gates, and a specific single qubit non-unitary gate, $\mathcal{N}_1(a)$, which relies on a single qubit projective measurement. $\mathcal{N}_1(a)$ has the matrix representation
\begin{equation}\label{eq:N1}
    \mathcal{N}_1(a) = \begin{pmatrix} 1 & 0 \\ 0 & a \end{pmatrix},
\end{equation}
for some $a\in[0,1]$. Its circuit construction is given in Figure \ref{fig:N1}; it is composed of a controlled single qubit unitary gate $\mathcal{U}_1(a)$, whose matrix representation is
\begin{equation}\label{eq:U1}
    \mathcal{U}_1(a) = \begin{pmatrix} a & * \\ * & * \end{pmatrix},
\end{equation}
along with a single ancillary measurement. Again, entries given by an asterisk $*$ indicate that there is some freedom in choosing their values; specifically, a single qubit unitary is described by three degrees of freedom. Note that $|a|\leq1$ is a necessary requirement for $\mathcal{U}_1$ to be unitary. \newline

Written explicitly using little endian ordering -- that is, in the $\ket{\phi_\text{anc}} \otimes \ket{\psi}$ basis, where $\ket{\phi_\text{anc}}$ denotes the ancillary qubit state -- the unitary corresponding to the circuit in Figure \ref{fig:N1}, $U_{\mathcal{N}_1}(a)$, is a block encoding of $\mathcal{N}_1$:
\begin{equation} 
    U_{\mathcal{N}_1}(a) = \begin{pmatrix} \textcolor{red}{1} & \textcolor{red}{0} & 0 & 0 \\ \textcolor{red}{0} & \textcolor{red}{a} & 0 & * \\ 0 & 0 & 1 & 0 \\ 0 & * & 0 & * \end{pmatrix} = \begin{pmatrix} \textcolor{red}{\mathcal{N}_1(a)} & * \\ * & * \end{pmatrix}. \label{eq:n1_isablockencoding}
\end{equation}


Upon post-selecting the 0 measurement on the ancilla qubit one obtains the $\mathcal{N}_1(a)$ matrix of elements shown in red:

\begin{equation}
    \big(|0 \rangle \langle 0 | \otimes I \big) U_{\mathcal{N}_1}(a) \big( |0 \rangle \langle 0 | \otimes I\big) = \mathcal{N}_1(a).
\end{equation}

Using this decomposition, successful implementation of a non-unitary circuit corresponds to a measurement shot in which all ancillary measurements from the constituent $\mathcal{N}_1(a)$ gates yield $\ket{0}$. \newline

\section{Trotterised PITE Circuit} \label{sec:method}
We consider building the ITE operator, $\tfrac{1}{\alpha}e^{-\hat{H}\tau}$, in terms of $\mathcal{N}_1(a)$ single qubit non-unitary gates \eqref{eq:N1}. Each constituent $\mathcal{N}_1(a)$ gate is implemented as a block encoding \eqref{eq:n1_isablockencoding} according to Figure \ref{fig:N1}, and the overall block encoding we achieve for the ITE operator will determine its scaling $\tfrac{1}{\alpha}$ (Section \ref{sec:block_encoding}). \newline

This is a natural approach to take when the ITE operator has been Trotterised. Given a Hamiltonian expressed in a qubit (Pauli) basis, the Trotter decomposition approximately factors the ITE operator into a product of exponentiated Pauli strings. Since each of these terms can be easily diagonalised with the use of single qubit rotations (refer to Section \ref{sec:trotter_decomposition}), they can be readily represented in terms of the diagonal $\mathcal{N}_1(a)$ matrices. \newline



The proof presented by Terashima et al \cite{terashima2005nonunitary} could be regarded as a general synthesis strategy for non-unitary operators; however, no attempt has been made to optimise the number of gates for the specific non-unitary operation of ITE. Moreover, the proof relies on an initial singular value decomposition (SVD) of the non-unitary matrix in question, incurring a classical cost that scales exponentially with the number of qubits $n$, $\mathcal{O}\left(2^{3n}\right)$. Instead, we consider a decomposition based on the Pauli gadget framework for Trotterised real time evolution (RTE) \cite{nielsenchuang2011}. This allows us to transfer any optimisation passes previously designed for real time propagation with Pauli gadgets directly to our framework \cite{cowtan2020phase}. \newline


\subsection{Trotter decomposition}\label{sec:trotter_decomposition}


Consider an $n$ qubit Hamiltonian that can be expressed as a sum of $M$ local interactions, $\hat{H} = \sum_{m=1}^{M} \hat{H}_m$, where $M \sim \mathcal{O}\left( \text{poly}(n) \right)$ and each Hamiltonian term is Hermitian. Applied to an imaginary time evolution $\tau = it$, product formulae approximate the evolution under the full Hamiltonian, $e^{-\hat{H}\Delta\tau}$, as a sequence of operators, $e^{- \hat{H}_m\Delta \tau}$, which should each be efficiently implementable on a quantum computer. For instance, the first-order Trotter formula states that \cite{lloyd1996universalsimulators}
\begin{equation}\label{eq:trotter_one}
    e^{-\hat{H}\Delta\tau} = \prod_{m=1}^{M} e^{- \hat{H}_m\Delta \tau} + \mathcal{O}((L \Delta \tau)^2),
\end{equation}
where $L$ is the system size. For a longer simulation time, the evolution is divided into $r \in \mathbb{N}$ Trotter steps

\begin{equation}\label{eq:trotter}
    e^{-\hat{H}\tau} = \left(\prod_{m=1}^{M} e^{- \hat{H}_m \Delta \tau}\right)^r + \mathcal{O}\left( (L \Delta \tau)^2 r \right),
\end{equation}

where $\frac{\tau}{r} = \Delta \tau$ and $r$ is also referred to as the `Trotter number'.
From Equation \eqref{eq:trotter}, we can determine the gate count $N$ required for a given evolution time $\tau$ and target precision $\epsilon$:
\begin{equation}\label{eq:trotter_error}
    \epsilon = \mathcal{O}(L^2 \tau \Delta \tau) \text{,  } N = Lr = \mathcal{O}\left(\frac{L^3\tau^2}{\epsilon}\right).
\end{equation}


The main challenge is to choose $r$ to be as small as possible and still ensure a total simulation error of at most $\epsilon$. This choice is complicated by the fact that the upper bound on the Trotter error in Equation \eqref{eq:trotter_error} can be rather loose. In particular, the error is strongly dependent on the commutator structure of the Hamiltonian; for instance, in the limiting case, all $M$ Hamiltonian terms commute and the error is exactly zero. Consideration of the commutator structure can be used to reduce the theoretical error scaling with respect to $L$ \cite{childs2021trottererror}; however, simulations can remain orders of magnitude faster than theory even when many Hamiltonian terms commute \cite{lloyd1996universalsimulators, babbush2015chemical, childs2018speedup}. \newline

Grouping terms into mutually commuting partitions can reduce the circuit depth significantly: not only is the Trotter error reduced, allowing for fewer Trotter steps to be taken, but the ordering of terms within a mutually commuting partition can be optimised -- with no effect on the Trotter error -- to maximise gate cancellations \cite{cowtan2020phase}. Partitioning the Hamiltonian into the minimum number of mutually commuting groups is equivalent to the minimum clique cover problem, which is NP-hard \cite{miller1972nphardpauligrouping}; thus, it is typically performed heuristically \cite{verteletskyi2020measurementoptimisation, huggins2021efficientmeasurements}. \newline

The product formula approach is highly flexible. For instance, higher order product formulae can be defined \cite{ostmeyer2023optimised}; the higher the order, the more accurate the Trotterisation. This presents a trade-off between the order of the decomposition and $r$: increasing the order reduces the number of Trotter steps $r$ required to achieve a fixed $\epsilon$, whilst also increasing the cost per Trotter step. Recent approaches to Trotter error mitigation have considered randomly permuting the Hamiltonian terms \cite{zhang2012randomised, childs2019fasterquantum}. Further, classical stochastic ITE (quantum Monte Carlo) methods randomly sample Hamiltonian terms, effectively weighted by the population of walkers located between connected basis states. They rely on the outcome that, in so doing, it is still possible to converge to the ground state \cite{booth2009fciqmc, spencer2012fciqmc, thom2010stochasticcc}. Similarly, the qDRIFT algorithm randomly samples Hamiltonian terms weighted by their coefficients: despite largely forsaking knowledge of the internal commutative structure of the Hamiltonian, it eliminates the explicit dependence of the gate count on the system size (Equation \eqref{eq:trotter_error}) \cite{campbell2019random, chen2021concentration, pocrnic2023composite, nakaji2023qswift}. In this paper, we only implement the first-order Trotter formula of Equation \eqref{eq:trotter_one} in our experimental simulations and theoretical examination, however, our techniques can be applied to any of the other decompositions. \newline


Given a fermionic Hamiltonian in second quantised form, the Hamiltonian must first be converted to a qubit representation. That is, fermionic creation and annihilation operators are mapped onto qubit operators (namely, tensor products of Pauli operators, or `Pauli strings'). To this end, we will use the Jordan-Wigner (JW) transformation \cite{jordan1928jw}, which maps electronic configurations onto computational basis states; for instance, the state $\ket{01}$ indicates a system with two spin orbitals, in which the first (second) spin orbital is unoccupied (occupied). Thus, for the remainder of this paper, we will consider Hamiltonians expressed in a Pauli basis,
\begin{equation}\label{eq:hamiltonian_local_decomposition}
    \hat{H} = \sum_{k=1}^{M} c_k \hat{\sigma}_{\boldsymbol{m}_k},
\end{equation}
where $\hat{\sigma}_{\boldsymbol{m}}$ is a Pauli string of length $n$,
\begin{equation}
    \hat{\sigma}_{\boldsymbol{m}} = \hat{\sigma}_{m_{n-1}} \otimes ... \otimes \hat{\sigma}_{m_0}, \; \; \; \hat{\sigma}_{m} \in \{I, X, Y, Z\}.
\end{equation}

\subsection{Pauli gadgets for Trotterised PITE} \label{sec:PG_for_PITE}

We now approach the task of block encoding the non-unitary operator, $\tfrac{1}{\alpha}e^{-\hat{H}\tau}$ for $\hat{H}=\hat{H}^\dag$, via a Trotter decomposition. Each Hamiltonian term $\hat{H}_m = \hat{H}_m^\dag$ in the Trotter decomposition can be diagonalised by a unitary operator $\hat{U}_{m}$, yielding:
\begin{equation} \label{eq:basis_transformation_general}
    e^{-\hat{H}_{m}\tau} = \hat{U}_{m} e^{-\hat{D}_m\tau} \hat{U}_{m}^\dag.
\end{equation}
In particular, Pauli strings are unitarily mapped onto other Pauli strings via Clifford operations. The Clifford group can be generated by three gates: the Hadamard gate ($H$), the phase gate ($S$), and the {\sc cnot} gate. We could choose to diagonalise each Pauli string individually:
\begin{equation} \label{eq:basis_transformation_ps}
    e^{-\hat{\sigma}_{\boldsymbol{m}_k}c_k\tau} = \hat{U}_{\boldsymbol{m}} e^{-\hat{\Lambda}_{\boldsymbol{m}_k}c_k\tau} \hat{U}_{\boldsymbol{m}}^\dag, \; \; \; \hat{U}_{\boldsymbol{m}} = \bigotimes_{k=0}^{n-1}\hat{B}_{m_{k}},
\end{equation}

where $\hat{\Lambda}_{\boldsymbol{m}_k} = \{Z, I\}^{\otimes n}$ and $\hat{B}_{m_k}$ denotes the basis transformation operator from the $\hat{\sigma}_{m_k}$-basis to the computational basis: $\hat{B}_{X} = H$, $\hat{B}_{Y} = SH$ and $\hat{B}_{I} = \hat{B}_{Z} = I$. Alternatively, we could simultaneously diagonalise a mutually commuting subset of Pauli strings; in this case, the diagonalisation can be constructed efficiently from Clifford operations in a classical preprocessing step \cite{vandenberg2020simultaneous, kawase2023simultaneous}. This approach can reduce the circuit depth significantly when large numbers of terms form a mutually commuting subset, since it allows for more {\sc cnot} gate cancellations. It is often also applied as a standard procedure for reducing the number of shots required for the evaluation of expectation values \cite{verteletskyi2020measurementoptimisation, huggins2021efficientmeasurements}. \newline

Pauli grouping methods are applicable to general qubit Hamiltonians. However, methods that produce a compressed representation of the original Hamiltonian typically yield better performance when applicable \cite{huggins2021efficientmeasurements}; until recently, the leading method for reducing the cost of Hamiltonian simulation for quantum chemistry Hamiltonians was tensor hyper contraction, which uses non-unitary rotations and therefore restricts the method to LCU approaches \cite{cohn2021hypercontractioncompare}. There are now superior double factorisation techniques which use unitary rotations \cite{oumarou2023accelerating} and can be combined with Trotter decompositions and its non-unitary variants, including the algorithm presented in this work. \newline

For simplicity, we will adopt the first approach (Equation \eqref{eq:basis_transformation_ps}) in this work. \newline



Once this change of basis has occurred, the exponentiated diagonal Pauli string must be implemented. Using the ubiquitous phase gadget circuit structure (refer to Appendix \ref{appendix:appendix_pauli_gadget}), this task requires us to find a block encoding for the single-qubit diagonal non-unitary operator $e^{+|\gamma| \Delta\tau \hat{Z}}$, for any $\gamma \in \mathbb{R}$. We note that a circuit structure implementation of the  operator $e^{+|\gamma| \Delta\tau (\hat{Z} - \hat{I})}$ has recently been proposed \cite{zhang2021observation}; our work can be seen as an extension of this to include any Hamiltonian expressed in a Pauli basis. \newline

As discussed in Section \ref{sec:block_encoding}, $\tfrac{1}{\alpha}$ is a scaling factor that depends on the block encoding used. The greater this scaling is, the greater the success probability of the block encoding; its maximum allowable value, which still ensures that the block encoding is unitary and thus implementable on a quantum circuit, is discussed in the following section (Section \ref{sec:max_alpha_ITE}). \newline


\subsubsection{Maximum scaling for the block encoded ITE operator, \texorpdfstring{$\left(\tfrac{1}{\alpha}\right)_{\text{max}} e^{-\hat{H}\tau}$}{TEXT}} \label{sec:max_alpha_ITE}

First, we consider the exact (non-Trotterised) ITE operator. According to Equation \eqref{eq:max_alpha}, the maximum possible scaling for the block-encoded exact ITE operator is given by
\begin{equation}
    \left(\frac{1}{\alpha}\right)_{\text{PITE, max}} = \frac{1}{\|e^{-H\tau}\|} = \frac{1}{\sigma_{\text{max}} \left(e^{-H\tau}\right)} = e^{E_0 \tau}, \label{eq:max_ITE_alpha}
\end{equation}
where $E_0$ is the (unknown) ground state energy. Once the ground state is reached, the block encoded ITE operator cannot project out any more states: continued application does not change the state, but if the scaling is less than the maximal value, the success probability will be less than 1, $p_{\text{success}} = \tfrac{1}{\alpha^2} e^{-2E_0}$. On the other hand, if the scaling is maximal, the ITE operator will be applied deterministically, $p_{\text{success}} = 1$. In other words, this maximal scaling ensures that the ITE operator becomes the identity operator once the ground state is reached. \newline

Note that this scaling is also present in our original definition of the ITE procedure for obtaining the ground state of a system \eqref{eq:ITE_operator}. When using ITE methods which do not iteratively renormalise the wavefunction, the prefactor in this definition ensures that the amplitude of the wavefunction does not decay to zero as the ground state is approached. Typically this will require re-scaling of the operator; consider, for instance, classical stochastic ITE methods (quantum Monte Carlo), in which the wavefunction is propagated in imaginary time by sampling the action of the Hamiltonian in some discrete basis populated by `walkers'. Since the total number of walkers at any one point in the evolution is related to the normalisation of the wavefunction, the ground state energy is estimated $\tilde{E}_0 \sim E_0$ throughout the run-time of the algorithm and used to shift the Hamiltonian, $e^{-\left(\hat{H} - \tilde{E}_0\right)\tau}$, as a means of walker population control \cite{booth2009fciqmc, spencer2012fciqmc, thom2010stochasticcc, filipucc2022}. On the other hand, PITE algorithms naturally renormalise the intermediate state $|\psi_i\rangle$ at each time step $i$ through the use of partial measurements:

\begin{equation}
    |\Psi\rangle = \prod_{i=1}^r \frac{e^{-\hat{H}\Delta\tau}}{\sqrt{\langle \psi_i| (e^{-\hat{H}\Delta\tau})^\dagger e^{-\hat{H}\Delta\tau} | \psi_i \rangle }}|\psi_0\rangle,
\end{equation}
    
rather than affecting the normalisation of the quantum state, due to the cancellation shown in Equation \ref{eq:renormalisation}. The scaling $\tfrac{1}{\alpha}$ affects only the success probability of these measurements (Section \ref{sec:block_encoding}). Thus, the factor of $e^{E_0 \tau}$ in Equation \eqref{eq:ITE_operator} is not required by PITE to prevent the decay of the ground state: $\tfrac{1}{\alpha}< e^{E_0 \tau}$ is allowed. Instead, this factor manifests as the optimal scaling $\left(\tfrac{1}{\alpha}\right)_{\text{max}}$ that minimises the decay of the success probability. We may want to take inspiration from quantum Monte Carlo methods and vary the scaling of the block encoding. However, since this is a non-unitary operation, it is more difficult to implement it using quantum processes. \newline

It is likely that PITE algorithms will need to be run in conjunction with an AA procedure to boost the success probability of the circuits. The PITE block encoding proposed by Liu et al \cite{liu2021pite} was defined in terms of an extra parameter; they were able to vary this parameter throughout the simulation, alongside Grover's algorithm \cite{grover1996fast, grover1997haystack}, to provide more flexible optimisation of the number of Grover iterates required. In a similar manner, Nishi et al \cite{nishi2022accelerating} were able to make $\tfrac{1}{\alpha}$ a tunable parameter within the LCU-based PITE framework \cite{kosugi2021probabilistic}, with reduced computational cost as compared with the fixed point search \cite{grover2005fixedpoint, yoder2014fixedpoint} or oblivious AA routine \cite{berry2014exponential}. As will become apparent, the value of $\tfrac{1}{\alpha}$ for the block encoding we propose in Section \ref{sec:block_encoding_trottITE} is initially fixed by the circuit implementation we use. However, we note that our framework is also amenable to varying the value of the scaling throughout the simulation, although as discussed, this is not a trivial task and does not lie within the scope of this work. \newline


Whilst it is easy to determine the maximum scaling for the exact ITE operator, it is not obvious how to relate this to the Trotterised form, which would depend on the Trotter error \cite{childs2021trottererror}. Weak upper and lower bounds for the maximum scaling of the Trotter decomposed operator, defined by Equations \eqref{eq:trotter} and \eqref{eq:hamiltonian_local_decomposition}, can be obtained (Appendix \ref{appendix:appendix_max_alpha}):
\begin{equation}
    e^{- \lambda r \Delta\tau} \leq \left(\frac{1}{\alpha}\right)_{\text{TrottPITE, max}} \leq e^{\lambda r \Delta\tau}, \label{eq:alpha_max_trottITE}
\end{equation}
where $\lambda = \sum_{k=1}^{M}|c_k|$ is the 1-norm of the Hamiltonian. Using the following relation between the spectral norm and the 1-norm of the Hamiltonian expressed in a Pauli basis:
\begin{equation}
    \|\hat{H}\| \leq \sum_{k=1}^{M} |c_k| \cdot \underbrace{\|\hat{\sigma}_{\boldsymbol{m}_{k}}\|}_{=1} = \lambda,
\end{equation}
we find that the maximum scaling for the exact ITE operator also lies in this range,
\begin{equation}
    e^{- \lambda r \Delta\tau} \leq e^{E_0 \tau} = \left(\frac{1}{\alpha}\right)_{\text{PITE, max}} \leq e^{\lambda r \Delta\tau}. \label{eq:alpha_max_ITE}
\end{equation}
\subsubsection{Block encoding for \texorpdfstring{$e^{-\gamma \Delta\tau \hat{Z}}$}{TEXT} with a single ancillary qubit}\label{sec:simple_ITE_block_encoding}

As a preliminary to block encoding the exponentiated Hamiltonian, we first consider the task of block encoding the simple operator $\tfrac{1}{\alpha}e^{-\gamma\Delta\tau \hat{Z}}$, where $\gamma \in \mathbb{R}$. According to Equation \eqref{eq:max_alpha}, this operator is implemented with the maximum possible success probability when the rescaling is given by:
\begin{equation}
    \left(\frac{1}{\alpha}\right)_{\text{max}} =\frac{1}{\|e^{-\gamma \Delta\tau Z}\|}=e^{- |\gamma|\Delta\tau }. \label{eq:simple_ITE_alpha}
\end{equation}
The matrix representation of the optimally block encoded operator $\left(\tfrac{1}{\alpha}\right)_{\text{max}} e^{-\gamma \Delta\tau \hat{Z}}$ can be easily expressed in terms of the $\mathcal{N}_1(a)$ matrix structure (Section \ref{sec:measurement_scheme}). Consider first the case $\gamma<0$:
\begin{equation}\label{eq:circuit_defn_neg}
    e^{-|\gamma|\Delta\tau}e^{-\gamma \Delta\tau Z} = \begin{pmatrix} 1 & 0 \\ 0 & e^{-2|\gamma|\Delta\tau} \end{pmatrix} = \mathcal{N}_1(e^{- 2|\gamma|\Delta\tau}).
\end{equation}

Similarly, for the case $\gamma>0$, we identify:
\begin{equation}\label{eq:circuit_defn_pos}
    e^{-|\gamma|\Delta\tau}e^{-\gamma \Delta\tau Z} = \begin{pmatrix} e^{-2|\gamma|\Delta\tau} & 0 \\ 0 & 1 \end{pmatrix} = X\mathcal{N}_1(e^{- 2|\gamma|\Delta\tau})X.
\end{equation}
In order to implement Equations \eqref{eq:circuit_defn_neg} and \eqref{eq:circuit_defn_pos} using the circuit structure defined in Figure \ref{fig:N1}, we must identify a single qubit unitary gate $\mathcal{U}_1(e^{-2|\gamma|\Delta\tau})$. Setting $\phi = 2\arccos{\left( e^{-2|\gamma|\Delta\tau} \right)}$, we find that we can easily implement $\mathcal{U}_1(e^{-2|\gamma|\Delta\tau})$ as a rotation, $R_x(\phi)$. Overall, the implementation of $\left(\tfrac{1}{\alpha}\right)_{\text{max}}\hat{A} = e^{-|\gamma|\Delta\tau}e^{-\gamma \Delta\tau \hat{Z}}$ is obtained with the circuits shown in Figures \ref{fig:PITE_primitive_combined_neg} and \ref{fig:PITE_primitive_combined_pos}, provided that the ancillary state is measured to be in the $\ket{0}$ state. \newline

\begin{figure}[!htb]
\centering
    \begin{adjustbox}{width=0.99\columnwidth}
    \begin{quantikz}
        \lstick[wires=1]{$\frac{1}{\sqrt{2}}\big(\ket{0} + \ket{1} \big)$} & \ctrl{1} & \qw \rstick[wires=1] {$\tfrac{1}{\sqrt{p_{\text{s}}}} \left(\ket{0} + e^{- 2|\gamma|\Delta\tau}\ket{1}\right)$} \\
        \lstick[wires=1]{$\ket{0}$} & \gate{R_x(\phi)} & \meterD{0}
    \end{quantikz}
    \end{adjustbox}

    \begin{adjustbox}{width=0.99\columnwidth}
    \begin{quantikz}[ampersand replacement=\&]
    \centering
       \lstick[wires=1]{$\frac{1}{\sqrt{2}}\big(|0\rangle + |1\rangle \big)$}  \& \gate{\left(\begin{array}{cc} 1 & 0 \\ 0 &  e^{- 2|\gamma|\Delta\tau} \end{array}\right)} \& \qw \rstick[wires=1] {$\tfrac{1}{\sqrt{p_{\text{s}}}}\left(\ket{0} + e^{- 2|\gamma|\Delta\tau}\ket{1} \right)$}
    \end{quantikz}
    \end{adjustbox}
\caption{Quantum circuit realisation of the non-unitary operator $e^{-|\gamma| \Delta\tau}e^{-\gamma \Delta\tau \hat{Z}}$ for $\gamma<0$ via post-selection on $\ket{0}$, where $\phi = 2\arccos{\left( e^{-2|\gamma|\Delta\tau} \right)}$. The post-selected state is renormalised by the square root of the success probability, $p_{\text{success}}$, denoted here $p_{\text{s}} = 1 + e^{-4|\gamma|\Delta\tau}$.}
\label{fig:PITE_primitive_combined_neg}
\end{figure}

\begin{figure}
    \begin{adjustbox}{width=0.99\columnwidth}
    \begin{quantikz}
        \lstick[wires=1]{$\frac{1}{\sqrt{2}}\big(\ket{0} + \ket{1} \big)$} & \octrl{1} & \qw \rstick[wires=1] {$\tfrac{1}{{\sqrt{p_{\text{s}}}}} \left(e^{- 2|\gamma|\Delta\tau}\ket{0} + \ket{1}\right)$} \\
        \lstick[wires=1]{$\ket{0}$} & \gate{R_x(\phi)} & \meterD{0}
    \end{quantikz}
    \end{adjustbox}
    
    \begin{adjustbox}{width=0.99\columnwidth}
    \begin{quantikz}[ampersand replacement=\&]
    \centering
       \lstick[wires=1]{$\frac{1}{\sqrt{2}}\big(|0\rangle + |1\rangle \big)$}  \& \gate{\left(\begin{array}{cc} e^{- 2|\gamma|\Delta\tau} & 0 \\ 0 &  1 \end{array}\right)} \& \qw \rstick[wires=1] {$\tfrac{1}{\sqrt{p_{\text{s}}}}\left(e^{- 2|\gamma|\Delta\tau} \ket{0} + \ket{1} \right)$}
    \end{quantikz}
    \end{adjustbox}
\caption{Quantum circuit realisation of the non-unitary operator $e^{-|\gamma| \Delta\tau}e^{-\gamma \Delta\tau \hat{Z}}$ for $\gamma>0$ via post-selection on $\ket{0}$, where $\phi = 2\arccos{\left( e^{-2|\gamma|\Delta\tau} \right)}$. The post-selected state is renormalised by the square root of the success probability, $p_{\text{success}}$, denoted here $p_{\text{s}} = 1 + e^{-4|\gamma|\Delta\tau}$.}
\label{fig:PITE_primitive_combined_pos}
\end{figure}

\FloatBarrier
\subsubsection{Block encoding for Trotterised \texorpdfstring{$e^{-\hat{H} \tau}$}{TEXT}} \label{sec:block_encoding_trottITE}

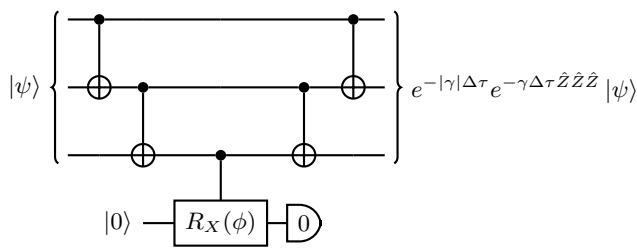
\begin{figure}[ht]
\centering
    \begin{quantikz}[column sep=0.25cm, row sep={0.9cm,between origins}]
        \lstick[wires=3]{$\ket{\psi}$}
         & \ctrl{1} & \qw & \qw & \qw & \ctrl{1}  & \qw \rstick[wires=3]{$e^{-|\gamma|\Delta\tau}e^{- \gamma \Delta\tau \hat{Z}\hat{Z}\hat{Z}}\ket{\psi}$} \\
        & \targ{} & \ctrl{1} & \qw & \ctrl{1} & \targ{}  & \qw \\
       & \qw & \targ{} & \ctrl{1} & \targ{} & \qw & \qw \\
        & & \lstick{$\ket{0}$} & \gate{R_X(\phi)} & \meterD{0} & & 
    \end{quantikz}
\label{fig:nonunitary_ZZZ_a}
\caption{Quantum circuit realisation of the non-unitary operator $e^{-|\gamma|\Delta\tau}e^{-\gamma\Delta\tau \hat{Z}\hat{Z}\hat{Z}}$, with $\gamma<0$ and $\phi = 2\arccos{\left( e^{-2|\gamma|\Delta\tau} \right)}$.}
\label{fig:nonunitary_ZZZ}
\end{figure}

Since the $\ket{0}$ post-selected two qubit gate becomes a diagonal non-unitary single qubit gate, we can use the same machinery of the $N$-qubit parity gate, otherwise known as the phase gadget (Figure \ref{fig:phase_gadget} in Appendix \ref{appendix:appendix_pauli_gadget}). Combining the phase gadget structure with the non-unitary single qubit gate (Figures \ref{fig:PITE_primitive_combined_neg} and \ref{fig:PITE_primitive_combined_pos}) gives the overall block encoding for the operator $e^{-|\gamma|\Delta\tau}e^{-\gamma\Delta\tau \hat{Z}^{\otimes n}}$ (Figure \ref{fig:nonunitary_ZZZ}). \newline

%




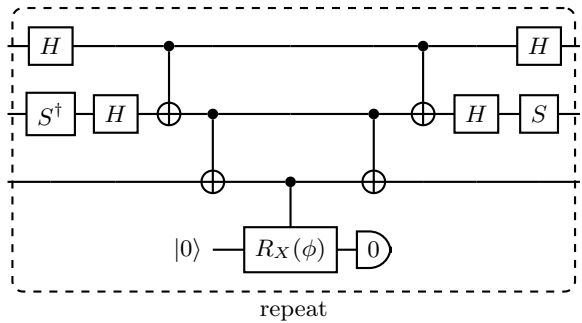
\begin{figure}
\centering
\sbox0{
\scalebox{1}{
\begin{quantikz}[column sep=0.25cm, row sep={0.9cm,between origins}]
    & \gate{H} \gategroup[4,steps=9,style={dashed, rounded corners, fill=white, inner xsep=2pt}, background, label style={label position=below, anchor=north, yshift=-0.2cm}]{repeat} & \qw & \ctrl{1} & \qw & \qw & \qw & \ctrl{1} & \qw & \gate{H} & \qw \\
    &\gate{S^\dag} & \gate{H} & \targ{} & \ctrl{1} & \qw & \ctrl{1} & \targ{} & \gate{H} & \gate{S} & \qw \\
    &\qw & \qw & \qw & \targ{} & \ctrl{1} & \targ{} & \qw & \qw & \qw & \qw \\
    & & & & \lstick{$\ket{0}$} & \gate{R_X(\phi)} & \meterD{0} & & & &
\end{quantikz}
}
}
\begin{tabular}{c}
\usebox0
\end{tabular}
\caption{Quantum circuit realisation of the non-unitary operator $e^{-|\gamma|\Delta\tau} e^{- \gamma \Delta\tau \hat{X}\hat{Y}\hat{Z}}$, with $\gamma<0$ and $\phi = 2\arccos{\left( e^{-2|\gamma|\Delta\tau} \right)}$.}
\label{fig:pauli_gadget_imag}
\end{figure}

An example circuit for the ITE of the Pauli string $\hat{X}\hat{Y}\hat{Z}$ is given in Figure \ref{fig:pauli_gadget_imag}; only a small modification needs to be made compared with the widely-used RTE Pauli gadget (Figure \ref{fig:pauli_gadget_real}). The overall circuit is then formed by concatenating circuits corresponding to each of the exponentiated Pauli strings, with their corresponding basis transformation gates \eqref{eq:basis_transformation}, and for each of the Trotter steps, according to the Trotter decomposition defined in Equations \eqref{eq:trotter} and \eqref{eq:hamiltonian_local_decomposition}. Since the ITE Pauli gadget for each Pauli string contains one $\mathcal{N}_1(e^{-|c_k| \Delta \tau \hat{Z}})$ block encoding circuit, which has a scaling of $e^{-|c_k| \Delta \tau}$ (Equation \eqref{eq:simple_ITE_alpha}), the scaling $\left(\frac{1}{\alpha}\right)_{\text{PITE $\hat{\sigma}_{\boldsymbol{m}_k}$}}$ for each Pauli gadget is also $e^{-|c_k| \Delta \tau}$ and is therefore maximised (Section \ref{sec:simple_ITE_block_encoding}):
\begin{equation}
    \left(\frac{1}{\alpha}\right)_{\text{PITE $\hat{\sigma}_{\boldsymbol{m}_k}$, max}} = \frac{1}{\|e^{-c_k \Delta \tau \sigma_{\boldsymbol{m}_k}}\|} = e^{-|c_k|\Delta \tau}.
\end{equation} \newline

From this, the overall scaling for the Trotterised PITE block encoding is given by
\begin{align}
    \begin{split}
    \left(\tfrac{1}{\alpha}\right)_{\text{TrottPITE}}
    &= \exp{-r \Delta \tau \sum_{\mathclap{\substack{k=1 \\ \hat{\sigma}_{\boldsymbol{m}_{k}} \neq \hat{I}^{\otimes n}}}}^{M} |c_{k}|} \label{eq:alpha_1_norm}\\
    &\geq \exp{-r\Delta \tau \lambda},
    \end{split}
\end{align}
where the sum in the second line \eqref{eq:success_prob_trott} is carried out over all coefficients of the constituent Pauli strings in $\hat{H}$, excluding any Pauli tensors of identity operators $\hat{I}^{\otimes n}$, and $\lambda = \sum_{k=1}^{M}|c_k|$ is the 1-norm of the Hamiltonian. Equality is achieved when the Hamiltonian does not include the term $\hat{I}^{\otimes n}$. Although each individual Trotter term has been optimally block encoded, the scaling for the overall Trotter product will depend on the order of the Trotter terms and may not be optimal (Equation \eqref{eq:alpha_max_trottITE}). \newline

Successful application of the Trotterised ITE operator requires all mid-circuit ancillary measurements to yield $\ket{0}$ - if a measurement is unsuccessful, the state is projected into the wrong subspace. Given a Trotter decomposition defined according to Equations \eqref{eq:trotter} and \eqref{eq:hamiltonian_local_decomposition}, the success probability after $r$ Trotter steps is given by
\begin{widetext}
\begin{align}
    \begin{split}
    p_{\text{success}} &= 
    \tfrac{1}{\alpha^2} \bra{\psi} A^\dag A \ket{\psi} \\
    &=\exp{-2r \Delta \tau \sum_{\mathclap{\substack{k=1 \\ \hat{\sigma}_{\boldsymbol{m}_{k}} \neq \hat{I}^{\otimes n}}}}^{M} |c_{k}|}
    \bra{\psi} \left(\left(\prod_{k=1}^{M} e^{- \hat{\sigma}_{\boldsymbol{m}_k}c_k \Delta \tau}  \right)^{\dag}\right)^{r} \left(\prod_{k=1}^{M} e^{- \hat{\sigma}_{\boldsymbol{m}_k} c_k \Delta \tau}\right)^{r} \ket{\psi}
    \end{split} \label{eq:success_prob_trott} \\
    &\geq \exp{-2r \Delta \tau \lambda}
    \bra{\psi} \left(\left(\prod_{k=1}^{M} e^{- \hat{\sigma}_{\boldsymbol{m}_k} c_k \Delta \tau}\right)^{\dag}\right)^{r} \left(\prod_{k=1}^{M} e^{- \hat{\sigma}_{\boldsymbol{m}_k} c_k \Delta \tau}\right)^{r} \ket{\psi} \label{eq:success_prob_trott_1norm}\\
    \begin{split}
    &\geq \exp{-2r \Delta \tau \lambda}
    \bra{\psi} \left(\prod_{k=1}^{M} e^{-|c_k| \Delta \tau} \right)^{r} \left(\prod_{k=1}^{M} e^{-|c_k| \Delta \tau} \right)^{r} \ket{\psi} \label{eq:success_prob_lower_bound} \\
    &= \exp{-4r \Delta \tau \lambda}. \end{split}
\end{align}
\end{widetext}


As expected, the success probability decays exponentially with the number of time steps taken and depends heavily on $\ket{\psi}$. The lowest success probability occurs for the state $\ket{\psi}_{\text{worst}}$ which undergoes pure amplitude damping under the action of every exponentiated operator in the Trotter decomposition -- that is, $e^{- \hat{\sigma}_{\boldsymbol{m}_k} c_k \Delta \tau} \ket{\psi}_{\text{worst}} = e^{-|c_k|\Delta\tau}\ket{\psi}_{\text{worst}}$ for all $k$, giving rise to the final inequality in Equation \eqref{eq:success_prob_lower_bound}. This would require $\ket{\psi}$ to be a simultaneous eigenstate of all the Pauli strings comprising $\hat{H}$, which is only possible when all the Pauli strings comprising $\hat{H}$ commute; however, consideration of this scenario does provide us with a lower bound for the general case. \newline

It is evident that the success probability for this algorithm can be increased by reducing the 1-norm of the Hamiltonian  \eqref{eq:success_prob_lower_bound}; this task is also of central importance to the reduction of gate complexity for several simulation algorithms, including simulation based on an LCU decomposition  \cite{childs2012lcu}. Loaiza et al \cite{loaiza2023reducing} have carried out significant work towards reducing the 1-norm of a molecular Hamiltonian; for instance, they describe several approaches for changing the Pauli basis decomposition of the Hamiltonian \eqref{eq:hamiltonian_local_decomposition}, including grouping together anti-commuting Pauli strings. When applied to the Trotterised PITE procedure, we must consider that whilst using these methods to reduce the 1-norm of the Hamiltonian would result in a higher success probability, it may also increase the Trotter error, resulting in a higher gate complexity. An alternative approach for reducing the 1-norm is to apply a symmetry shift operator to the Hamiltonian, $e^{-\hat{H}\tau}=e^{-\hat{S}\tau}e^{-(\hat{H}-\hat{S})\tau}$, such that the 1-norm of the operator $\hat{H}-\hat{S}$ is less than that of $\hat{H}$. Provided the state vector satisfies symmetry constraints, for real time evolution $t = -i\tau$, the unitary operator $e^{-i\hat{S}t}$ applies a phase shift to the state vector and can be ignored. However, this cannot be exploited for imaginary time evolution: shifting the Hamiltonian leaves behind a non-unitary operator, $e^{-\hat{S}\tau}$, which, as previously discussed in Section \ref{sec:max_alpha_ITE}, is non-trivial to implement as a quantum circuit. \newline





In practice, currently one must run a number of circuits to the end of the simulation, and the shots for which all ancillary measurements yielded $\ket{0}$ -- that is, the successful shots -- are post-selected. Since the probability of successful application decreases exponentially with the number of mid-circuit measurements, the number of successful shots from which to estimate the expectation value also decreases exponentially; a sufficiently high number of shots must be selected to ensure that the final energy is determined with high accuracy. In future applications, the implementation of a quit--if--fail functionality would significantly improve the overall run-time of the algorithm. \newline

The gate complexity and the total number of shots required to obtain an accurate estimate of the ground state energy at the end of the ITE procedure depend on the total evolution time $\tau$ for convergence. Consider expressing an arbitrary state $\ket{\psi(0)}$ in the eigenbasis of $\hat{H}$, $\ket{\psi(0)} = \sum_{i=0}^{2^n-1} a_i \ket{\phi_i}$, where $\hat{H}\ket{\phi_i} = E_i \ket{\phi_i}$ and $E_0<E_1<...<E_{2^n-1}$ \cite{jin2022quantum}. The unnormalised state after a time $\tau$ is
\begin{equation}
    \ket{\psi(\tau)} = \sum_{i=0}^{2^n-1} a_i e^{-E_i \tau}\ket{\phi_i},
\end{equation}

and its normalisation is given by
\begin{align}
    \begin{split}
    \mynorm{\ket{\psi(\tau)}}^2 &= 
    \sum_{i=0}^{2^n-1} |a_i|^2 e^{-2E_i \tau} \\
    &= |a_0|^2 e^{-2E_0\tau}\left(1 + \sum_{i>0}^{2^n-1} \left(\tfrac{|a_i|}{|a_0|}\right)^2 e^{-2(E_i-E_0) \tau}\right).
    \end{split}
\end{align}
To specify how long we need to evolve the state, let $\tau$ be the time required to achieve an accuracy of $\epsilon \ll 1$ in the squared overlap with the ground state,
\begin{equation}
    \frac{| \bra{\psi(\tau)}\ket{\phi_0} |^2}{\mynorm{\ket{\psi(\tau)}}^2} \geq 1-\epsilon.
\end{equation}
In the large $\tau$ limit, only the ground state and first excited state contributions are left:
\begin{equation}
    \frac{|a_0|^2 e^{-2E_0\tau}}{|a_0|^2 e^{-2E_0\tau}\left(1 + \left(\tfrac{|a_1|}{|a_0|}\right)^2 e^{-2(E_1-E_0) \tau}\right)} \geq 1-\epsilon.
\end{equation}
\begin{equation}
    1 + \left(\frac{|a_1|}{|a_0|}\right)^2 e^{-2(E_1-E_0) \tau} \lesssim \frac{1}{1-\epsilon}.
\end{equation}
The number of times, $r_{\epsilon} = \tfrac{\tau}{\Delta \tau}$, that a time step of $\Delta \tau$ must be applied to achieve an error, $\epsilon$, is thus given by
\begin{align}
    r_{\epsilon} &\gtrsim \frac{1}{2(E_1 - E_0)\Delta \tau} \ln{\left( \frac{|\alpha_1|^2}{\epsilon |\alpha_0|^2} \right)}. \label{eq:n_Trotter}
\end{align}
As expected, the more easily distinguishable the ground state and first excited state energies are, the faster convergence to the ground state is achieved. Further, the better the initial guess, i.e. the greater the initial overlap with the ground state, the faster convergence is achieved. \newline


Combining Equations \eqref{eq:success_prob_lower_bound} and \eqref{eq:n_Trotter}, assuming $\Delta \tau$ is small enough to be able to neglect the Trotter error, would in principle give a lower bound on the total number of shots required to obtain an error of $\epsilon$ in the ground state estimation. \newline

The circuit depth required for convergence to the GS is then determined by the number of Pauli gadgets, $rM$, where $M$ is the number of local terms comprising the Hamiltonian \eqref{eq:hamiltonian_local_decomposition}. Each of these Pauli gadgets requires $2n$ {\sc cnot} gates; thus each Pauli gadget produces a gate depth that scales at worst linearly in $n$, and at best logarithmically in $n$ \cite{cowtan2020phase}. The whole procedure can be implemented using one ancillary qubit by resetting it to $\ket{0}$ following every mid-circuit measurement. 


\section{Results} \label{sec:results}

Pytket \cite{sivarajah2020pytket} is used for the construction and compilation of the circuits, and all quantum simulations are performed with the Qiskit Aer simulator \cite{treinish2022qiskit}. For each of the Hamiltonians, the constituent Pauli strings are partitioned into mutually commuting sets to reduce the Trotter error. \newline



Energies and their associated errors are estimated in the following manner. Individual Pauli strings are sampled and the mean energy is constructed as
\begin{equation}
    \langle \hat{H} \rangle = \sum_{k=1}^{M} c_k \langle \hat{\sigma}_{\boldsymbol{m}_{k}} \rangle.
\end{equation}

The error $\Delta E$ in the mean energy arising from using a finite number of shots is determined using a similar method to Kandala et al \cite{kandala2017hardware}:
\begin{equation}
    \Delta E = \sqrt{\sum_{k=1}^{M} \frac{c_k^2 \langle \Delta \hat{\sigma}_{\boldsymbol{m}_{k}}^2 \rangle}{N_{k}}},
\end{equation}

where $\langle \Delta \hat{\sigma}_{\boldsymbol{m}_{k}}^2 \rangle$ is the variance on Pauli string $\hat{\sigma}_{\boldsymbol{m}_{k}}$ and $N_k$ is the number of successful shots used in the measurement of $\langle \Delta \hat{\sigma}_{\boldsymbol{m}_{k}}^2 \rangle$.

\subsection{Models studied}

\subsubsection{Ising Model}
The transverse field Ising model (TIM) \cite{pfeuty1970tim} is the simplest spin model that reveals interesting properties of quantum magnetism, such as quantum phase transitions and quantum spin glasses, and can been used to simulate quantum annealing \cite{kadowaki1998tim}. In this paper, given the na\"ive implementation of the algorithm, we restrict ourselves to the one-dimensional case, for which the spectrum can be found analytically by a JW transformation from the spin model onto free fermions \cite{pfeuty1970tim} (note that this is the reverse of the JW mapping mentioned elsewhere in this work, which refers to encoding fermionic degrees of freedom into qubits). \newline



The Hamiltonian for the TIM with $n$ sites and periodic boundary conditions is defined as:
\begin{equation}\label{eq:ising}
    \hat{H} = -J\sum_{i=1}^{n} \hat{Z}_i \hat{Z}_{i+1} - h \sum_{i=1}^{n} \hat{X}_i,
\end{equation}

where $\hat{Z}_{n+1} = \hat{Z}_{1}$. The Ising model with $n$ sites is represented with $n$ qubits. Under the JW transformation, the TIM Hamiltonian is composed of $2n-1$ Pauli strings. \newline


We prepare the initial state to be an equal superposition of all spin basis states using the Hadamard gate: $H^{\otimes n} \ket{0}^{\otimes n}$. This represents a uniform prior -- that is, we have not encoded any prior information about what we might expect the ground state (GS) to be. Importantly, since we consider the ferromagnetic limit, $J>h$, we do not benefit from a good initial guess for the GS.


\subsubsection{Hubbard model}




The fermionic Hubbard model (FHM) is the simplest possible model for correlated electrons; it approximates long-range Coulomb interactions with on-site interactions, but still exhibits a wide range of interesting phenomena, including magnetic ordering, metal-insulator transition and superconductivity \cite{leblanc20152dhubbard, qinhubbardcomputationalperspective}. Further, the FHM exhibits correlations that are difficult to capture by classical methods. To this end, the Hubbard model is widely used as a benchmark for quantum algorithms. Again, we restrict our analysis to the one-dimensional chain under periodic boundary conditions, for which an analytic solution is known \cite{lieb1968exact1dhubbard}. \newline





The Hamiltonian for the non-relativistic single-band fermionic Hubbard model in real space is given by:



\begin{equation}\label{eq:hubbard}
    \hat{H} = t\sum_{\lambda = \uparrow, \downarrow}\sum_{i, j} \hat{c}_{i, \lambda}^{\dag} \hat{c}_{j, \lambda} + U\sum_{i} \hat{c}_{i, \uparrow}^{\dag} \hat{c}_{i, \uparrow} \hat{c}_{i, \downarrow}^{\dag} \hat{c}_{i, \downarrow},
\end{equation}

where $U>0$ corresponds to repulsive on-site electron-electron interactions, and $t<0$ corresponds to a lowering of the kinetic energy of the system by allowing for delocalisation over the sites. We consider the case where electrons are strongly interacting -- that is, when $\left|\frac{U}{t}\right| \geq 1$. The sum over $i, j$ is typically restricted to account for the exponentially decaying overlap of wavefunctions between sites. We adopt the standard model, in which hopping is only considered between nearest-neighbours. \newline

The Hubbard model with $m$ sites is represented by $2m$ qubits under the JW transformation. An occupation number basis is used to enumerate the states in the Hilbert space; for instance, the 2-site model has basis states $\ket{n_{1 \uparrow} n_{1 \downarrow} n_{2 \uparrow} n_{2 \downarrow}}$. The Hamiltonian under the JW transformation is comprised of $7m-3$ Pauli strings, including $\hat{I}^{\otimes 2m}$.\newline

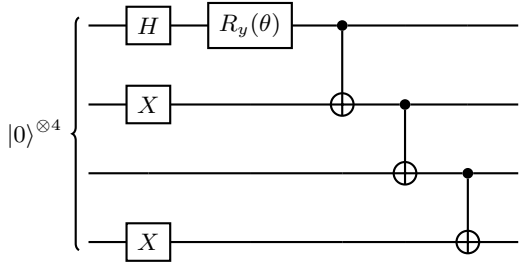
\begin{figure}[h!]
\centering
\sbox0{
\begin{quantikz}
    \lstick[wires=4]{$\ket{0}^{\otimes 4}$}& \gate{H} & \gate{R_y(\theta)} & \ctrl{1} & \qw  & \qw & \qw\\
    & \gate{X} & \qw & \targ{} & \ctrl{1} & \qw & \qw\\
    & \qw & \qw & \qw & \targ{} & \ctrl{1} & \qw \\
    & \gate{X} & \qw & \qw & \qw & \targ{} & \qw
\end{quantikz}
}
\begin{tabular}{c}
 \\
\usebox0
\end{tabular}
\caption{Initial state preparation circuit for the 2 site 1D Hubbard model. This circuit produces the state $\tfrac{1}{\sqrt{2}} (\cos{\tfrac{\theta}{2}}-\sin{\tfrac{\theta}{2}}) \ket{0110} + \tfrac{1}{\sqrt{2}} (\cos{\tfrac{\theta}{2}}+\sin{\tfrac{\theta}{2}}) \ket{1001}$. When $\theta = \pi$, the singlet state is prepared, $\tfrac{1}{\sqrt{2}}(\ket{0110} - \ket{1001})$. This circuit structure can be easily generalised to generate a linear combination of the two classically antiferromagnetic states for any number of sites.}
\label{fig:initial_state_prep_hubbard}
\end{figure}

Since the Hubbard Hamiltonian conserves the total number of spin up and spin down electrons, $[\hat{H}, \hat{n}_{\uparrow}] = [\hat{H}, \hat{n}_{\downarrow}] = 0$, we can consider the action of the Hamiltonian on a particular sector $(n_{\uparrow}, n_{\downarrow})$ of Hilbert space \cite{hubbardnotes}. For a system of $m$ sites, we consider the half-filled sector $(m/2, m/2)$. Figure \ref{fig:initial_state_prep_hubbard} gives the initial state preparation circuit; the X and {\sc cnot} gates are used to excite the state into the  sector $(m/2, m/2)$, whereas the H and $\text{R}_y$ gates produce a superposition of the two antiferromagnetic states: $\tfrac{1}{\sqrt{2}}(\cos{\tfrac{\theta}{2}}-\sin{\tfrac{\theta}{2}}) \ket{011001...} + \tfrac{1}{\sqrt{2}}(\cos{\tfrac{\theta}{2}}+\sin{\tfrac{\theta}{2}}) \ket{100110...}$. We choose $\theta = \pi$, giving an out-of-phase superposition of the two states. \newline

\subsection{Simulation Results}
\begin{figure*}
    \centering
    \subfloat{\includegraphics[width=0.66\columnwidth]{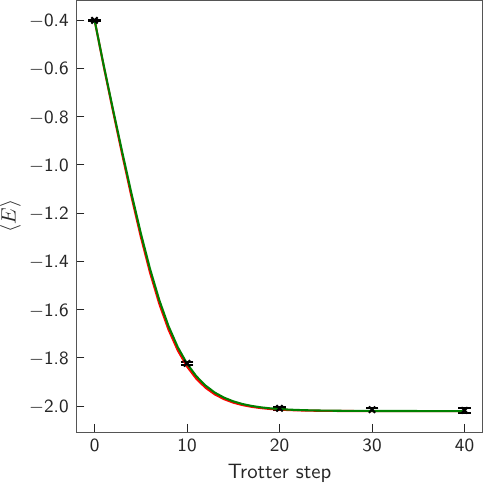}}\,
    \subfloat{\includegraphics[width=0.66\columnwidth]{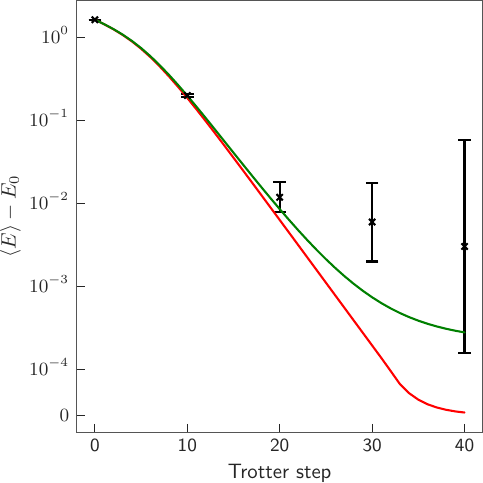}}\,
    \subfloat{\includegraphics[width=0.66\columnwidth]{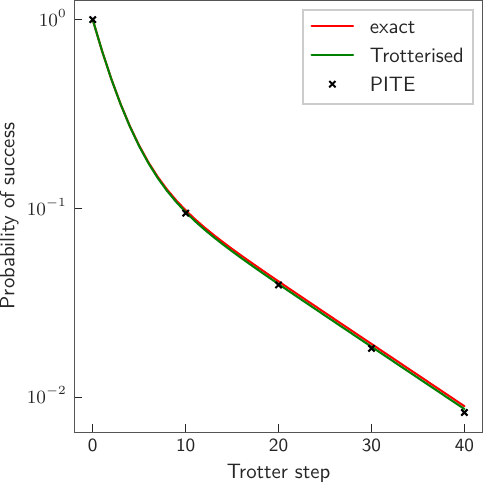}}
    \caption{Energy estimate (left panel), relative energy compared to the true GS energy, $E_0$ (centre panel) and success probability (right panel) plotted as a function of the number of time steps for the 4-site TIM Hamiltonian, with $J=0.5$, $h=0.1$ \eqref{eq:ising}, and $\Delta \tau = 0.1$. The red curves are the exact results obtained from an exact diagonalisation of the Hamiltonian, the green curves are the expected results from the Trotterised ITE operator, and the black points have been evaluated with the Trotterised PITE algorithm using $10^5$ measurement shots.}
    \label{fig:4siteTIM_time}
\end{figure*}

\begin{figure*}
    \centering
    \subfloat{\includegraphics[width=0.66\columnwidth]{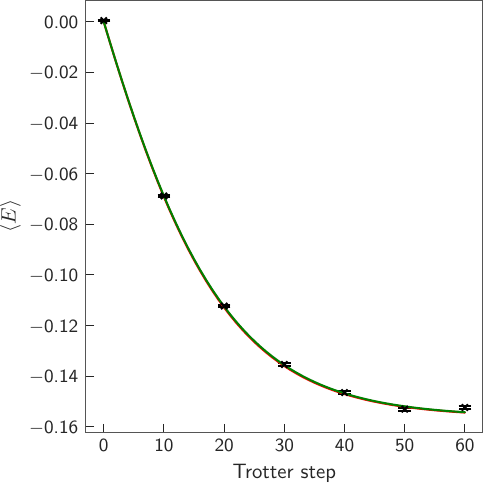}}\,
    \subfloat{\includegraphics[width=0.66\columnwidth]{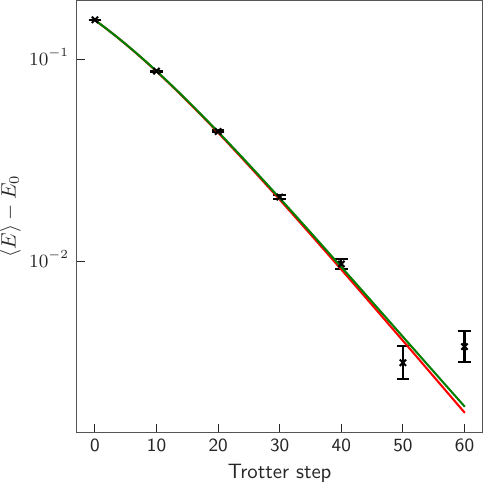}}\,
    \subfloat{\includegraphics[width=0.66\columnwidth]{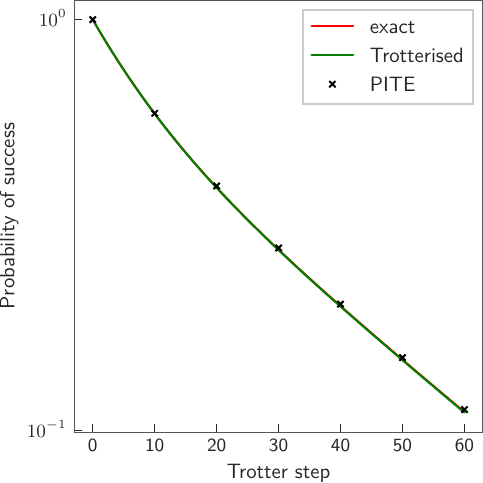}}
    \caption{Energy estimate (left panel), relative energy compared to the true GS energy, $E_0$ (centre panel) and success probability (right panel) plotted as a function of the number of time steps for the 2-site Hubbard Hamiltonian, with $t=-0.1$, $U=0.1$ \eqref{eq:hubbard}, and $\Delta \tau = 0.1$. The red curves are the exact results obtained from an exact diagonalisation of the Hamiltonian, the green curves are the expected results from the Trotterised ITE operator, and the black points have been evaluated with the Trotterised PITE algorithm using $10^5$ measurement shots.}
    \label{fig:2sitehubbard_time}
\end{figure*}

Figures \ref{fig:4siteTIM_time} and \ref{fig:2sitehubbard_time} show the simulation results for the 4-site TIM and the 2-site Hubbard model respectively; both the expected energy and the fraction of successful measurement shots are plotted and compared to the exact results, obtained from an exact diagonalisation of the Hamiltonian, as well as to the expected evolution of the Trotterised ITE operator. There is good agreement between the expected Trotter results and the simulation results; deviations not accounted for by the error calculations can be attributed to additional stochastic errors. \newline

We observe the expected exponential decay in the total probability of successful application – the longer the evolution time required for convergence to the ground state, the fewer the number of successful shots remaining that can be used in the final estimation of the energy. As expected, the stochastic errors in the mean energy increase throughout the evolution as the number of successful shots decreases. As the system size increases, the final success probability at convergence is reduced. After increasing the number of sites in the Hubbard model to 3 and 4 sites, the success probability at convergence decreases to $10^{-5}$ and $10^{-9}$ respectively, requiring a prohibitive number of shots (roughly $10^9$ and $10^{13}$) to obtain an accurate estimation of the ground state. This being said, we also observe an initial exponential approach towards the GS energy. \newline

\section{Discussion} \label{sec:discussion}

The PITE algorithm presented in this paper, Trotterised PITE, is shown to systematically recover the ground state for the small systems investigated, giving incentive for further investigation. The most comparable PITE proposal to the method presented in this paper uses a block encoding constructed from a LCU \cite{kosugi2021probabilistic}. Whilst Trotterised PITE block encodes the Trotter-decomposed ITE operator, LCU PITE block encodes a first-order approximation to the ITE operator, $(1-\hat{H}\tfrac{\tau}{r})^r$, where $r$ is the number of time steps -- importantly, this is still a ground state projector in the limit $\tau \rightarrow \infty$ \cite{zhang2016wallch}. Similar to Trotterised PITE, the LCU algorithm uses one ancillary qubit, which is reset after each mid-circuit measurement, and unlike many of the previous PITE proposals, both methods automatically build the required circuits given any input Hamiltonian expressed in a Pauli basis. Throughout this discussion, we will make comparisons with the LCU PITE method. \newline

The cost of the PITE algorithm arises from two separate sources: (1) the exponential decay in the success probability with the evolution time, necessitating the use of large numbers of shots, from which the successful shots are post-selected, and (2) the depth of the circuits. The first of the two is the limiting factor, as is the case with most block encoding methods. Indeed, the simulation results demonstrate that, given an initial na\"ive implementation of the Trotterised PITE algorithm, it is prohibitively expensive to apply it to larger systems; for the Hubbard Hamiltonian, it was even infeasible to run simulations for more than 2 sites. For the purposes of GS determination, we only require that repeated application of the operator eventually projects the state onto the GS. Many GS projectors are possible \cite{zhang2016wallch}. Roughly speaking, the `better' the GS projector -- that is, the fewer the number of applications of the projector, $r$, required to reach the GS -- the more `information' is thrown away after each application, and the more rapidly the success probability decays. The results in Figures \ref{fig:4siteTIM_time} and \ref{fig:2sitehubbard_time} suggest that, though the success probability for the ITE operator decays exponentially with simulation time, there is an initial exponential approach towards the GS, suggesting that PITE methods could be used as part of a scalable routine for GS determination. For instance, PITE could be used for a shorter length of time, and the resulting state could then be passed to the QCELS method, which only requires a squared overlap with the GS of 0.5 \cite{ding2023qcels, ding2023qcelsbetter}. This routine would make use of the fast initial decay in excited state contributions from the exponential GS projector, whilst avoiding extra reduction in success probability in the latter, slower part of the convergence. \newline

Since all PITE proposals aim to approximate the ITE operator, they will all produce similar success probabilities at the end of the evolution, the only difference arising from the scaling factors $\tfrac{1}{\alpha}$ of these block encodings. This is because the success probability at any given time is only determined by the degree of non-unitarity of the operator -- that is, the success probability of a block encoding of the ITE operator is largely determined by the Hamiltonian, and to a lesser extent by the prefactor $\tfrac{1}{\alpha}$, which reduces the success probability by a factor of $\alpha^2$ \eqref{eq:success_probability}. The proposed Trotterised PITE method is optimal with respect to the scaling factor for each Hamiltonian term in the product formula. For the exact ITE operator, the scaling factor is not optimal but can be increased by reducing the 1-norm of the qubit Hamiltonian, a task that is of interest to several Hamiltonian simulation procedures; to this end, significant work has already been carried out \cite{loaiza2023reducing}. \newline

The most significant improvements to the success probability, however, would likely be achieved by running the PITE algorithm in tandem with some form of AA procedure \cite{liu2021pite, nishi2022accelerating}. It is important to appreciate that the design of this AA procedure is a non-trivial task and should be tailored to the specific PITE procedure in order to minimise circuit depths; for instance, Grover's algorithm was implemented with good success by Nishi et al for the LCU PITE proposal \cite{nishi2022accelerating}. \newline

The final success probability in a PITE procedure largely depends on the evolution time required for convergence to the ground state $\tau$, which is a property inherent to the system, since it is fixed by the energy difference between the ground state and first excited state of the system (Equation \ref{eq:n_Trotter}). However, $\tau$ is also dependent on the overlap between the initial guess wavefunction and the GS. When using a Jordan-Wigner transformation, which maps each spin orbital onto a qubit, subsequent entanglement of the qubits will encode a Full Configuration Interaction (FCI) wavefunction. It is important to note that, since this work presents a proof of principle implementation of a Trotterised PITE procedure, the initial states used in this work were particularly poor. Typically, the Hartree-Fock (HF) state is used as the initial guess: as the system size grows, the contribution of the HF state to the ground state diminishes \cite{szaboostlund1996advanced}, and $\tau$ must correspondingly increase. Indeed, for strongly correlated systems, there are simulations which suggest that the overlap decreases exponentially with system size \cite{lee2023exponentialqa}. Reducing $\tau$ requires choosing an initial state with a larger overlap with the GS: for instance, by running an initial classical simulation to screen for the most important amplitudes in the FCI expansion. Filip et al. used this approach to screen for important amplitudes in a Unitary Coupled Cluster Ansatz, used in a variational quantum eigensolver routine \cite{filipucc2022}. On quantum hardware, we have the additional complexity of needing to find methods of efficiently preparing the desired initial state \cite{sugisaki2019qcqc}. \newline

The mitigation of the exponential decay in the success probability is the most important obstacle for the implementation of PITE methods. However, we should also consider methods for reducing the circuit depths. Aside from being proportional to $\tau$, the minimum number of time steps required is also inversely related to the maximal time step $\Delta \tau$ that will still guarantee convergence to the ground state. This is, in turn, determined by the error in the approximation used. Trotterisation is a very popular simulation method due to its simplicity, flexibility and efficiency; as such, it is immediately possible to reduce the circuit depths of the Trotterised PITE method by appealing to the wealth of techniques already developed by the Trotter decomposition community. For instance, applying higher order decompositions \cite{ostmeyer2023optimised} and randomisation protocols \cite{childs2019fasterquantum, campbell2019random, chen2021concentration, pocrnic2023composite, nakaji2023qswift}, grouping together Hamiltonian terms to minimise the number of mutually commuting partitions \cite{babbush2015chemical, childs2018speedup, lloyd1996universalsimulators, vandenberg2020simultaneous, kawase2023simultaneous}, and reordering terms to maximise the number of gate cancellations \cite{cowtan2020phase}. Further, until recently, the leading method for reducing the cost of Hamiltonian simulation for quantum chemistry Hamiltonians was tensor hyper contraction, which uses non-unitary rotations and therefore restricts the method to LCU approaches \cite{cohn2021hypercontractioncompare}. There are now superior double factorisation techniques which use unitary rotations \cite{oumarou2023accelerating} and can be combined with Trotter decompositions and its non-unitary variants, including Trotterised PITE. \newline

Trotterised PITE produces shorter circuits than LCU PITE, at the cost of requiring more mid-circuit measurements. LCU PITE uses $2M$ controlled Pauli gadgets per time step, where $M$ is the number of Pauli strings comprising the Hamiltonian, followed by a single mid-circuit measurement. On the other hand, Trotterised PITE uses $M$ (not controlled) Pauli gadgets per time step, along with $M$ mid-circuit measurements. Since these Pauli gadgets are not controlled, they can be executed with higher fidelity. Note that the requirement for more mid-circuit measurements in our algorithm is not a restriction for some quantum computer architectures, like those based on trapped ions, as they are able to execute mid-circuit measurements on a similar timescale as gate operations. \newline

As mentioned, for the purpose of GS determination, the operator only needs to be a GS projector. However, for the purpose of computing thermal averages, it is important to be able to approximate the ITE operator to high accuracy \cite{tazhigulov2022correlated, motta2020determining}. LCU PITE is a block encoding of a first-order approximation to the ITE operator. Thus, regardless of the accuracy with which the constituent controlled RTE operations are performed, the dominant error will scale as $\left( \Delta \tau \right)^2$ per time step. On the other hand, applying the above methods to the Trotterised PITE method can not only reduce circuit depth, but can also be used to improve the overall error in the approximation to the ITE operator. \newline

The number of Hamiltonian terms may be minimised with a sensible choice of basis. Whilst we have used a second quantised basis throughout this paper, giving rise to a qubit representation of the Hamiltonian comprised of Pauli strings, it has been argued that a first quantised basis may produce more efficient scalings \cite{kosugi2021probabilistic}. Any ITE algorithm using a Pauli representation for the Hamiltonian can easily be extended to the first quantised basis using the protocol proposed by Kosugi et al. \cite{kosugi2021probabilistic, kosugi2022exhaustive}. \newline


Aside from improving the success probability, and thus reducing the number of shots required, we should also consider reducing the time taken for each shot to run. We note that improvements in hardware to provide a quit-if-fail functionality would remove the need to run failed shots to the end of the circuit, greatly reducing the average run-time across shots. \newline

\section{Conclusions} \label{sec:conclusion}


We have developed a purely quantum routine for performing probabilistic imaginary time evolution (PITE), based on a Trotter decomposition of the Hamiltonian. The block encoding suggested can be thought of as a modification of the Pauli gadget primitive, which is an efficient and widely used circuit implementation for real time evolution. PITE algorithms avoid many of the limitations of near-term algorithms. Namely, they avoid the restriction on the accuracy that can be achieved as a consequence of using a fixed ansatz in variational methods, including VITE, and any restrictions placed on the locality of the Hamiltonian, as is the case in QITE. \newline



In this paper, we have implemented the Trotterised PITE block encoding with a simple initial protocol: successful shots are post-selected, no amplitude amplification procedure is used, and minimal optimisation for the number of mid-circuit measurements required, and the depth of the circuits, is used. We applied this routine to the task of ground state determination in one-dimensional for the Transverse Ising Model with 4 sites and the fermionic Hubbard model with 2 sites. In particular, we found that the number of shots required by this na\"ive implementation became prohibitively high even for a 4-site fermionic Hubbard model. \newline

 We argue that this behaviour is expected given the nature of the PITE approach. The limiting factor in the performance of all PITE algorithms is that they exhibit an exponential decay in the probability of successful application with the number of mid-circuit measurements applied -- indeed, this problem is shared by many block encoding methods. Importantly, the algorithm successfully recovers the ground state for small systems, giving incentive to adapt its structure to overcome this limitation. We discussed a multitude of strategies that could be applied to reduce the run-time of the algorithm, of which the inclusion of an amplitude amplification procedure is likely to be the most significant contribution \cite{liu2021pite, nishi2022accelerating}. Further to this, Trotterised PITE could be used to initially increase the overlap with the ground state as part of a more scalable routine for ground state determination: (1) Trotterised PITE would first be run for a shorter length of time, during which many of the excited state contributions decay, (2) the resulting state is then passed to a quantum subspace diagonalisation method \cite{motta2020determining} or the QCELS algorithm, which only requires a squared overlap with the GS of $0.5$ \cite{ding2023qcels, ding2023qcelsbetter}. This routine would make use of the fast initial convergence of the exponential GS propagator, whilst avoiding unnecessary reduction in success probability in the latter, slower part of the convergence. \newline



ITE methods implemented on quantum circuits have found many applications in recent years. Sokolov et al used ITE for the determination of the ground state of transcorrelated Hamiltonians \cite{sokolov2022transcorrelated}. They obtained promising results, gaining up to four orders of magnitude improvement in the absolute energy error in comparison to non-transcorrelated approaches. More generally, algorithms that perform imaginary time evolution can be used as a subroutine; for instance, in the estimation of low-lying excited states using quantum subspace diagonalisation methods \cite{motta2020determining}. The most comparable method to Trotterised PITE is LCU PITE, which block encodes a first order approximation to the ITE operator. Since Trotterised PITE is a block encoding of the Trotter-decomposed ITE operator, whose accuracy can be readily improved, it is more suited than LCU PITE to applications which require an accurate application of the ITE operator; for instance, in the calculation of finite temperature correlation functions \cite{tazhigulov2022correlated, motta2020determining}. Moreover, we note that our algorithm can be readily extended to the real (or imaginary) time Trotterised simulation of non-Hermitian systems, $\hat{H} = \hat{H}_1 + i\hat{H}_2$, where $\hat{H}_1$ and $\hat{H}_2$ are Hermitian, by concatenating Pauli gadgets for real time propagation (deterministic, Figure \ref{fig:pauli_gadget_real}) and for imaginary time propagation (probabilistic, Figure \ref{fig:pauli_gadget_imag}). This could be used for the simulation of open systems (see, for instance, Algorithm I in \cite{kamakari2022opensystems}). \newline

\section*{Acknowledgements} \label{sec:acknowledgements}

The authors would like to thank Michael Foss-Feig and Gabriel Greene-Diniz for feedback on the manuscript.

\bibliographystyle{packages/apsrev4-2.bst}
\bibliography{references}

\onecolumngrid
\appendix*
\section{}
\subsection{Pauli gadgets for RTE}\label{appendix:appendix_pauli_gadget}

The following describes the Pauli gadget circuit primitive, which has been adopted by a wide range of quantum circuits. Given some analytic function, $f$, and some parameter, $t$, the operator $f(t \hat{Z}^{\otimes n})$ is diagonal in the computational basis. Thus, although $f(t \hat{Z}^{\otimes n})$ involves all the qubits in the system, it does so in a classical manner:
\begin{equation}\label{eq:analytic_phase}
    f\left(t\hat{Z}^{\otimes n}\right) = \sum_{i \in \{0,1\}^{n}} f(tp_i) |i\rangle \langle i|,
\end{equation}

where $p_i = +1$ if computational basis state $\ket{i}$ has even parity -- that is, the bitstring $i$ contains an even number of 1's -- and $-1$ if its parity is odd. \newline

\begin{figure}[h]
    \centering
    \begin{tikzpicture}
\node[scale=1] {
\begin{quantikz}[]
    \lstick[wires=3]{$\ket{\psi}$}
     & \ctrl{1} & \qw & \qw & \qw & \ctrl{1}  & \qw \rstick[wires=3]{$ZZZ\ket{\psi}$} \\
    & \targ{} & \ctrl{1} & \qw & \ctrl{1} & \targ{}  & \qw \\
   & \qw & \targ{} & \gate{Z} & \targ{} & \qw & \qw \\
\end{quantikz}
};
\end{tikzpicture}
    \caption{Phase gadget quantum circuit realisation of $ZZZ|\psi\rangle$.}
    \label{fig:phase_gadget}
\end{figure}
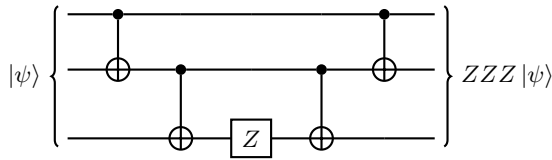

The simplest operator of this form is $\hat{Z}^{\otimes n}$. Figure \ref{fig:phase_gadget} exemplifies a quantum circuit realisation of this operator; circuits of this form are commonly known as `phase gadgets' \cite{nielsenchuang2011}. In this circuit, the `ladder of {\sc cnot}s' computes the parity of each of the computational basis states and encodes it in the state of the final qubit, setting its state to be $\ket{0}$ for an even parity and $\ket{1}$ for an odd parity. A $Z$-gate is then applied to this qubit, imparting a phase of $+1$ for an even parity and $-1$ for an odd parity, as required. Finally, the parity computation is reversed. \newline

The phase gadget primitive is widely used to simulate the real time evolution (RTE) of a Trotter-decomposed Hamiltonian. The RTE operator of a Pauli string $\hat{\sigma}_{\boldsymbol{m}}$ can be expressed as:
\begin{equation} \label{eq:basis_transformation}
    e^{-it\hat{\sigma}_{\boldsymbol{m}}} = \hat{U}_{\boldsymbol{m}} e^{-i \hat{Z}^{\otimes n} t} \hat{U}_{\boldsymbol{m}}^\dag, \; \; \; \hat{U}_{\boldsymbol{m}} = \bigotimes_{k=0}^{n-1}\hat{B}_{m_{k}},
\end{equation}

where $\hat{B}_{m_k}$ denotes the basis transformation operator from the $\hat{\sigma}_{m_k}$-basis to the computational basis. For instance, $\hat{B}_{X} = H$ and $\hat{B}_{Y} = SH$, where $H$ and $S$ denote the Hadamard gate and the phase gate, respectively. \newline

Equation \eqref{eq:basis_transformation} allows us to reduce the problem of simulating any Pauli string to the problem of simulating $\hat{Z}^{\otimes n}$. Once the phase gadget primitive is combined with the necessary basis transformation gates, the circuit is referred to as a `Pauli gadget'. According to Equation \eqref{eq:analytic_phase}, the effect of the RTE operator $e^{-it\hat{Z}^{\otimes n}}$ on each computational basis state is to add a phase conditioned on its parity, $e^{-itp_i}$. This can be achieved by modifying Figure \ref{fig:phase_gadget} to apply a $Z$ rotation gate, $R_z(2t)$, on the parity-storing qubit. \newline

\begin{figure}[h!]
\centering
\sbox0{
\scalebox{1}{
\begin{quantikz}[column sep=0.25cm, row sep={0.9cm,between origins}]
    & \gate{H} \gategroup[3,steps=9,style={dashed, rounded corners, fill=white, inner xsep=2pt}, background, label style={label position=below, anchor=north, yshift=-0.2cm}]{repeat} & \qw & \ctrl{1} & \qw & \qw & \qw & \ctrl{1} & \qw & \gate{H} & \qw \\
    &\gate{S^\dag} & \gate{H} & \targ{} & \ctrl{1} & \qw & \ctrl{1} & \targ{} & \gate{H} & \gate{S} & \qw \\
    &\qw & \qw & \qw & \targ{} & \gate{R_z(2\Delta t)} & \targ{} & \qw & \qw & \qw & \qw
\end{quantikz}
}
}
\begin{tabular}{c}
\usebox0
\end{tabular}
\caption{Quantum circuit realisation of $U(t) = e^{-i \Delta t XYZ}$.}
\label{fig:pauli_gadget_real}
\end{figure}
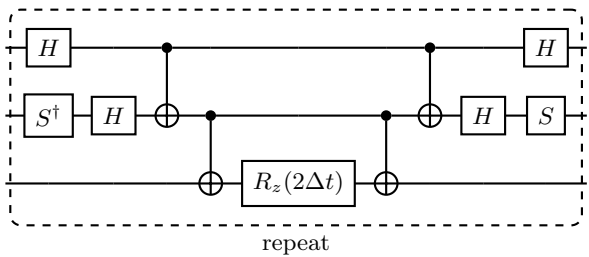

An example quantum circuit for the real time evolution of the Pauli string $XYZ$ is shown in Figure \ref{fig:pauli_gadget_real}. 
 The Pauli gadget is repeated for each of the Pauli strings, using their corresponding basis transformation gates \eqref{eq:basis_transformation}, and for each of the Trotter steps, according to the Trotter decomposition \eqref{eq:trotter}.


\subsection{Maximum scaling for the Trotterised ITE block encoding}\label{appendix:appendix_max_alpha}

Denoting $\sigma_{\text{max}}(B)$ and $\sigma_{\text{min}}(B)$ the maximum and minimum singular values of $B$, respectively, $\left(\tfrac{1}{\alpha}\right)_{\text{TrottPITE, max}}$ can be lower-bounded using the sub-multiplicative property of operator norms, $\|BD\| \leq \|B\| \cdot \|D\| = \sigma_{\text{max}}(B) \sigma_{\text{max}}(D)$:
\begin{align}
    \|A\| = \left\| \left( \prod_{k=1}^{M} e^{- \hat{\sigma}_{\boldsymbol{m}_k} c_k\Delta \tau} \right)^r \right\|
    &\leq \left(\prod_{k=1}^{M} \left\| e^{- \hat{\sigma}_{\boldsymbol{m}_k} c_k\Delta \tau} \right\| \right)^r \\
    &= \left(\prod_{k=1}^{M} e^{+ |c_k|\Delta \tau}\right)^r \\
    &= \exp{r\Delta \tau \lambda},
\end{align}
where $\lambda = \sum_{k=1}^{M}|c_k|$ is the 1-norm of the Hamiltonian. Similarly, we can achieve an upper bound using the inequality $\|BD\| \geq \sigma_{\text{min}}(B) \sigma_{\text{min}} (D)$:
\begin{align}
    \|\hat{A}\| = \left\| \left( \prod_{k=1}^{M} e^{- \hat{\sigma}_{\boldsymbol{m}_k} c_k\Delta \tau} \right)^r \right\|
    &\geq \left(\prod_{k=1}^{M}\sigma_{\text{min}}\left(e^{- \hat{\sigma}_{\boldsymbol{m}_k} c_k\Delta \tau} \right) \right)^r \\
    &= \left(\prod_{k=1}^{M} e^{- |c_k|\Delta \tau}\right)^r \\
    &= \exp{- r\Delta \tau \lambda}.
\end{align}

\end{document}